\documentclass[%
 reprint,
superscriptaddress,
 amsmath,amssymb,
 aps,
pre,
]{revtex4-2}

\usepackage{epsfig,dsfont,amssymb,amsmath,amsthm,amsfonts,amsbsy,mathrsfs}
\usepackage{graphicx}
\usepackage{color}
\usepackage{bm}
\usepackage{multirow}
\usepackage{natbib}
\usepackage[]{mdframed}

\newcommand{\nn}{\nonumber }

\newcommand{\BEQ}{\begin{equation}}
\newcommand{\EEQ}{\end{equation}}
\newcommand{\BEA}{\begin{eqnarray}}
\newcommand{\EEA}{\end{eqnarray}}

\newcommand{\blue}[1]{\textcolor{black}{#1}}

\usepackage{tikz}
\usetikzlibrary{arrows,shapes,chains}

\begin{document}
\preprint{APS/123-QED}
\title{\blue{Wave front propagation} in the Active Coagulation Model}

\author{Matteo Paoluzzi}
\email{matteo.paoluzzi@uniroma1.it}
\affiliation{Department of Physics, Sapienza University of Rome, Piazzale Aldo Moro, Rome, Italy.}

\date{\today}

\begin{abstract}
Spreading processes on top of active dynamics provide a novel theoretical framework for capturing emerging collective behavior in living systems. 
I consider run-and-tumble dynamics coupled with coagulation/decoagulation reactions that lead to an absorbing state phase transition. While the active dynamics does not change the location of the transition point, the relaxation toward the stationary state depends on motility parameters. Because of the competition between spreading dynamics and active motion, the system can support long-living currents whose typical time scale is a nontrivial function of motility and reaction rates. 
\blue{Because of this interplay between time-scales, the wave front propagation qualitatively changes from traveling to diffusive waves.}
\blue{Moving} beyond the mean-field regime, instability at finite length scales regulates a crossover from periodic to diffusive modes. Finally, it is possible to individuate different mechanisms of pattern formation on a large time scale, ranging from the Fisher-Kolmogorov to the Kardar-Parisi-Zhang equation. 
\end{abstract}

\maketitle
\section{Introduction}
\blue{In the biological world, most of the interactions do not have a strong constraint for being symmetric.}
Examples include social interactions \cite{fruchart2021non}, the synaptic dynamics in neural nets \cite{PhysRevLett.61.259}, \blue{and} biochemical reactions \cite{hinrichsen2000non}.  
Breaking symmetric interaction rules is another way to fall out of equilibrium \cite{PhysRevE.85.061127}. 
Recent studies have focused on non-symmetric interactions, particularly at the mesoscopic level \cite{PhysRevX.10.041009,pisegna2024emergent}, in both numerical simulations and coarse-graining theory \cite{dinelli2023non}, as well as minimal mixed spin models with distinct interaction rules for different spin variables \cite{avni2024dynamical}. 
\blue{The present work instead} focuses on the simplest scenario, familiar in statistical physics: reaction rules that are inherently non-symmetrical and typically lead to absorbing states, which break detailed balance \cite{hinrichsen2000non}. \blue{A typical application of absorbing state phase transitions is the population growth of a single species, where, in the absorbing state, the population becomes extinct while, in the mixed state, the balance between birth and death processes fixes the typical population size. The birth/date processes lead to non-linear equations that, once coupled with the diffusion process, can support wave propagations. Diffusion-driven (and thus noise-driven) wave propagation plays a crucial role in many biological processes (see \cite{murray2007mathematical} for details). The situation is much less clear if the population rearranges itself because of active dynamics. 
This is a case of practical interest: for instance, when two microbial bacterial strains compete with each other, segregation patterns emerge \cite{hallatschek2007genetic}, or more generally, in the case of chemical waves that play an important role in developmental biology \cite{deneke2018chemical}.}
This scenario naturally connects to the broader question of how spreading processes (e.g., \blue{Susceptible-Infected-Susceptible} 
or \blue{Susceptible-Infected-Removed} dynamics \cite{murray2007mathematical}), are affected by active dynamics.  
\blue{This is an emerging research field where the key question is what collective behavior results from the competition/cooperation of}
self-propelled motion and contagious dynamics 
\cite{PhysRevLett.100.168103,peruani2019reaction,PhysRevLett.133.058301,PhysRevResearch.4.043160,PhysRevResearch.2.032056,paoluzzi2020information,PhysRevE.98.052603,marcolongo2024assessing,peruani2013fluctuations,norambuena2020understanding,de2023sequential,PhysRevE.107.024604,forgacs2023transient}.

Here, \blue{I consider a minimal population growth model, the so-called coagulation model.} 
The coagulation model describes a single species undergoing coagulation and decoagulation reactions, which change the number of particles. 
These reactions introduce additional time scales that compete with those of the active dynamics.
Once we add self-propulsion, we can think of the model as a model of self-propelled agents that annihilate at a rate $\beta$ and spontaneously duplicate at a rate $\mu$. 
\blue{I will focus mostly on the mixed state, where the population size is set but the ratio $\mu/\beta$.}
While other active matter models \blue{in which the number of particles is not a conserved quantity}
have been explored \blue{in connection with pattern formation} \cite{cates2010arrested,curatolo2020cooperative}, this work focuses on how the competition of time scales \blue{produces qualitative different regimes in the front wave propagation of the active particle systems.}
This work shows that assuming rapidly decaying currents as a closure for the density field's time evolution is problematic when multiple time scales are present. \blue{This is because} long-living currents arise from the density-current coupling, where the coagulation rate is the coupling constant. When this coupling is not negligible, the system requires increasingly longer times to reach a stationary state. This increased relaxation time is associated with damped traveling waves that do not always relax monotonically to the stationary state.

\begin{figure}[!t]
\centering\includegraphics[width=.45\textwidth]{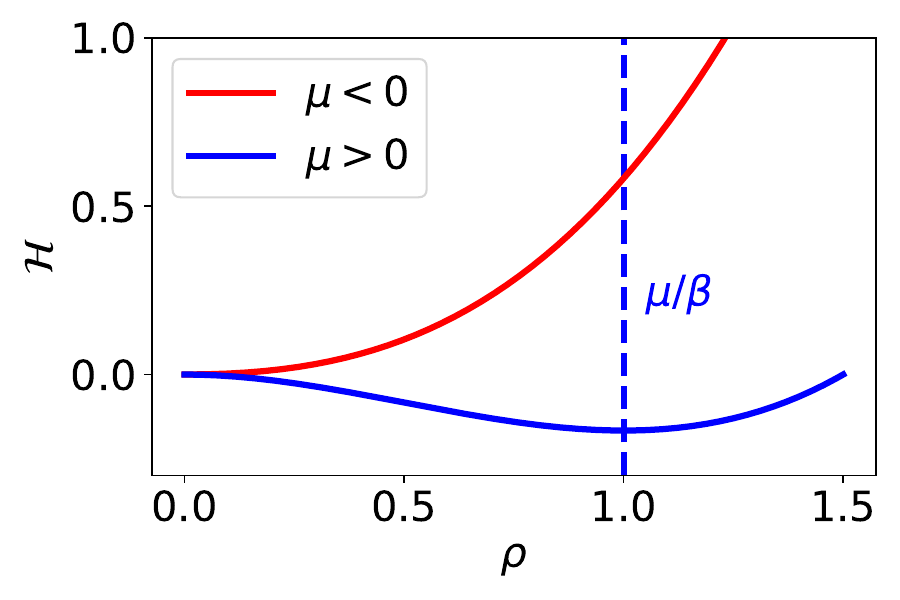}
\caption{Absorbing state phase transition. 
The function $\mathcal{H}$ (see (\ref{eq:H})) develops a minimum in $\rho=0$ for $\mu \leq 0$ that changes to $\rho = \mu / \beta \neq 0$
for $\mu>0$.}
\label{fig:model}      
\end{figure}

\section{Active Coagulation Model}
\begin{figure}[!t]
\centering\includegraphics[width=.45\textwidth]{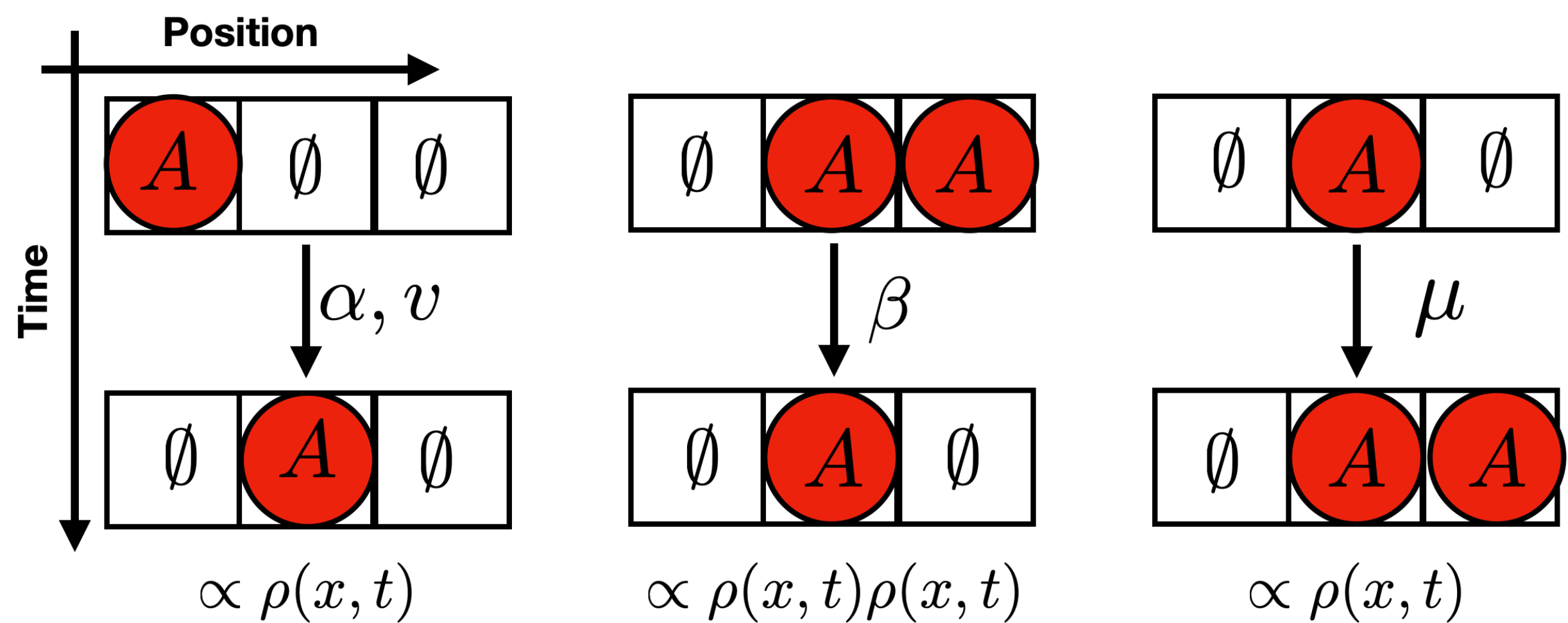}
\caption{\blue{Active coagulation model.  The active coagulation model involves three key processes: a run-and-tumble dynamics characterized by a tumbling rate $\alpha$ and self-propulsion velocity $v$, a coagulation process governed by the reaction rate $\beta$, and a decoagulation process controlled by the rate $\mu$. These processes are depicted in the top row as a lattice-gas model. Moving to a coarse-grained description, active motion and decoagulation contribute terms proportional to the density field $\rho(x,t)$, while the coagulation process introduces a nonlinear interaction term proportional to $\rho^2(x,t)$.}
}
\label{fig:model2}      
\end{figure}

\blue{To provide a general framework for studying the impact of active motion on population dynamics, I employ the so-called coagulation model that considers just two types
of reaction processes: a particle disappears with a density-dependent rate $\tilde{\beta}(\rho)$ and appears at a constant rate $\mu$, where $\rho$ represents the number density of active particles. The reaction process that makes particles disappear is called coagulation, and the reason why its rate is density-dependent will be clear later. The other reaction that allows particles to appear is called decoagulation. On top of that, I consider active motion within the so-called run-and-tumble dynamics. Without loss of generality, I will focus on the one-dimensional case that is general enough to capture the salient phenomenology of active matter. In run-and-tumble dynamics, we have two species of particles: the ones moving to the right and the ones moving to the left. I will consider both species moving at constant velocity. Right-moving particles invert their direction (in one spatial dimension, there is not enough space for a rotation of a finite angle) at a rate $\alpha$, usually called the tumbling rate. The same happens for the left-moving ones.}
Adopting the standard notation for run-and-tumble active particles in one spatial dimension, $R\equiv R(x,t)$ and $L\equiv L(x,t)$
indicate the fraction of right-moving and left-moving particles, respectively \cite{PhysRevE.48.2553}. The active motion is characterized by the self-propulsion velocity $v$ and the tumbling rate $\alpha$. In the limit $v \to \infty$ and $\alpha \to \infty$ with fixed diffusion constant $D_A = v^2 / \alpha$, the random walk reduces to the standard Brownian motion.
On top of the active motion, I consider a coagulation process characterized by two parameters. In the following, $\beta$ indicates the reaction rate of the coagulation reaction \blue{(that is now on a density-independent rate)}, and with $\mu$ the rate of the offspring production (decoagulation). 
\blue{The microscopic picture is the following.}
If $A$ indicates the presence of an active particle in some point of space at a given time, and $\emptyset$ indicates no particles, within coagulation dynamics one has to consider two elementary processes: the coagulation process that tends to annihilate particles, i.e, a particle spontaneously disappears, and the decoagulation process which introduces a new particle. 
After adding self-propelled motion, 
the set of reactions is
\begin{eqnarray}
&&A \emptyset \xrightarrow{\alpha, v} A \emptyset, \;\; \text{RT Dynamics} \\ \nn
&&A A \xrightarrow{\beta} A\emptyset, \;\; \text{Coagulation with rate} \; \beta \\ \nn
&&A \emptyset \xrightarrow{\mu} A A, \;\; \text{Decoagulation with rate} \; \mu \; .
\end{eqnarray}
\blue{The set of reactions is shown in Fig. (\ref{fig:model2}), which provides a pictorial representation of a microscopic lattice-gas version 
of the model under consideration. However, instead of considering the lattice-based picture, I will consider a suitable coarse-grained 
version obtained by noticing that coagulation processes are proportional to $\rho^2(x,t)$ while decoagulation is proportional to $\rho(x,t)$, with $\rho(x,t)$ the density field \cite{hinrichsen2000non}. In the case of run-and-tumble dynamics, on top of that, we have other "two species" of particles: the ones moving to the right and the ones moving to the left.}
I will employ these three elementary processes in the following fashion 
\begin{align}
    \dot{R} &= - v R^\prime - \frac{\alpha}{2} (R - L) - \beta R (R + L) + \mu R \\ 
    \dot{L} &= v L^\prime + \frac{\alpha}{2} (R - L) - \beta L (R + L) + \mu L
\end{align}
once one introduces the current $J(x,t) \equiv J = v(R - L)$ and the probability density $\rho(x,t) \equiv \rho = R + L$, the equations of motion for $J$ and $\rho$ are
\begin{align} \label{eq:rho}
    \dot{\rho} &= -J^\prime -\frac{\partial \mathcal{H}}{\partial \rho} \\ 
    \dot{J} &= -v^2 \rho^\prime - J \left[ \alpha + \mu - \beta \rho \right] \label{eq:J} \\ \label{eq:H}
    \mathcal{H}(\rho) &\equiv -\frac{\mu}{2} \rho^2 + \frac{\beta}{3} \rho^3 \; . 
\end{align}
The quadratic term \blue{in $\mathcal{H}$} generates linear interactions that tend to bring the system to extinction. Non-linear interactions \blue{with coupling constant $\beta$} are therefore essential for the existence of stationary solutions with $\rho \neq 0$.
The nonlinearities in (\ref{eq:rho}) generally render the dynamics analytically intractable.
Fig. (\ref{fig:model}) depicts the typical behavior of $\mathcal{H}$ as the mass term $\mu$ changes from negative to positive values. Since $\mu$ represents a rate, negative values are physically meaningless\blue{, however, in the spirit of critical phenomena where the mass tunes the distance from the critical point, $\mu$ should be understood in the same way as distance from the critical point where the two fixed points of the dynamics become a marginal point}. \blue{In any case, from the point of view of the coarse-grained model considered here, $\mu$ can take either positive or negative values.}

It is worth recalling the picture without coagulation dynamics, where the equations for $\rho$ and $J$ become
\begin{align} \label{eq:rt_standard_rho}
    \dot{\rho} &= -J^\prime \\ \label{eq:rt_standard_J}
    \dot{J} &= -v^2 \rho^\prime - \alpha J \; .
\end{align}
In this case, the dynamics of the system is governed by a second-order differential equation that can be obtained simply by computing $\ddot{\rho} = -\dot{J}^\prime$ so that
\begin{align}
    \ddot{\rho} + \alpha \dot{\rho} - v^2 \rho^{\prime\prime} = 0 \; .
\end{align}
This telegrapher equation predicts wave propagation on short-time scales that eventually diffuse \cite{kac1974stochastic}. One can rationalize that studying the limiting case $\alpha \to \infty$ at fixed diffusive constant $D_A \equiv v^2 / \alpha$ that reproduce the standard diffusive equation
\begin{align}
    \dot{\rho} = D_A \rho^{\prime \prime}
\end{align}
while in the opposite limit $\alpha \to 0$ one gets the standard wave equation
\begin{align}
    \ddot{\rho} = v^{2} \rho^{\prime \prime} \; ,
\end{align}
so that the model interpolates between a ballistic motion on time scales $t < \alpha^{-1}$ and a diffusive regime for $t > \alpha^{-1}$. The presence of an additional dynamical process, as in the case of the coagulation dynamics considered here, introduces another time scale that enters in competition with $\alpha^{-1}$. 
We also notice that, if we look at solutions of (\ref{eq:rt_standard_rho}) and (\ref{eq:rt_standard_J}) that do not depend on space, i.e., mean-field like solutions $\rho(x,t) = \rho(t)$ and $J(x,t) = J(t)$, one has $\dot{\rho} = 0$ and $\dot{J} = -\alpha J$, i.e., $\rho$ and $J$ decouple from each other, the current decays on the time scale $\alpha^{-1}$ ($J(t) = J(0) e^{-\alpha t}$ and the density is uniform $\rho(t) = \rho(0) = const.$

\begin{figure*}[!t]
\centering\includegraphics[width=1.\textwidth]{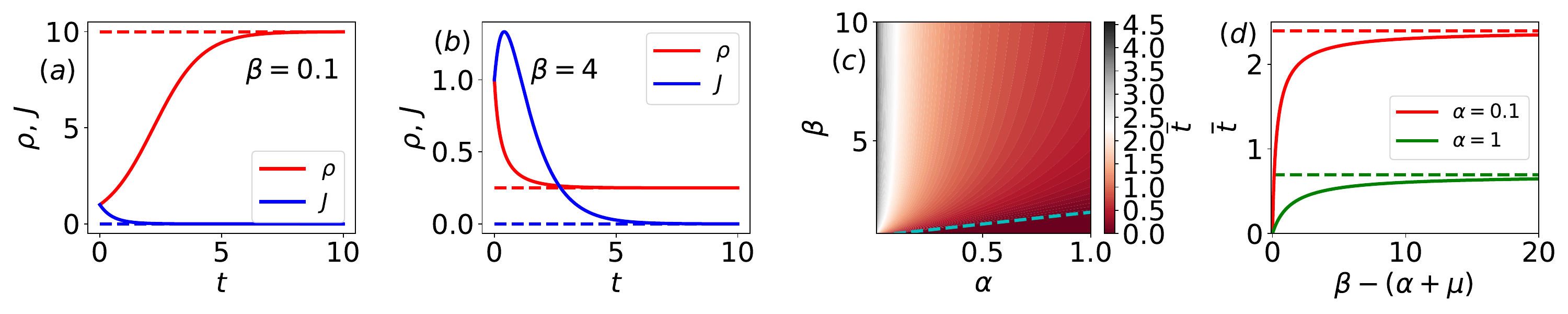}
\caption{Mean-Field approximation. (a) Time evolution of density $\rho$ and current $J$ for $\beta=0.1$ ($\alpha=\mu=v=1$). The dashed lines indicate the stationary values. (b) The current $J$ develops a peak at $\overline{t}$ for increasing values of $\beta$ (here $\beta=4$ and $\alpha=\mu=v=1$). (c) Contour plot $\overline{t}(\alpha,\beta)$. Below the dashed line $\overline{t}=0$. (d) $\overline{t}$ as a function of $\beta$ for $\alpha=0.1,1$ (see legend). The dashed lines indicate the asymptotic values $\beta \to \infty$.}
\label{fig:MF}      
\end{figure*}

\subsection*{Well-Mixed Approximation}
To understand what is the effect of an additional time scale,
I first study the mean-field picture that one obtains by neglecting any spatial dependency of $\rho$ and $J$ so that $\rho^\prime=J^\prime=0$. In this limit, one can find the analytical solution of the dynamics that turns out to be different from the standard RT dynamics where $\rho(t)$ is constant and $J(t)$ does not depend on $\rho(t)$ (see (\ref{eq:rt_standard_rho})). 
The equation for $\rho$ is the well-known logistic equation
\begin{align}
    \dot{\rho} = \mu \rho ( 1 - \frac{\beta}{\mu} \rho) 
\end{align}
with the analytical solution (with the initial condition 
$\rho(0)=1$)
\begin{align} \label{eq:dyn_rhoMF}
    \rho(t) = \frac{\mu}{\beta -(\beta - \mu) e^{-\mu t}} \; .
\end{align}
However, in contrast with a coagulation model in the mean-field approximation, in this case, there is still the equation for the current $J(t)$ that has to be taken into account. For the current $J(t)$ one gets
(with $\mu>0$)
\begin{align}
    \dot{J} = -J \left( \alpha + \mu - \beta \rho \right)
\end{align}
whose solution reads
\begin{align} \label{eq:dyn_JMF}
    J(t) = J(0) \, e^{- \int_0^t ds \, (\alpha + \mu - \beta \rho(s))} \; ,
\end{align}
once one plugs (\ref{eq:dyn_rhoMF}) into (\ref{eq:dyn_JMF}) with initial condition $J(0)=1$ it follows
\begin{align}
    J(t) &= \frac{G(t)}{G(0)} e^{-(\alpha + \mu) t} \\
    G(t) &\equiv \beta \left( 1 - e^{\mu t} \right) - \mu \; .
\end{align}
Figs. (\ref{fig:MF}a,\ref{fig:MF}b) report the typical behavior of $\rho(t)$ and $J(t)$.
Because of the persistent motion, one obtains non-trivial dynamics of the current $J$ that develops a peak before decaying to zero (see Fig. (\ref{fig:MF}b)). In particular, the exponential decay rate of the current, that is $\alpha$ in the case of RT dynamics, increases to $\alpha + \mu$ thanks to birth processes that produce at the same rate left-handed and right-handed particles. However, the coagulation process controlled by $\beta$ tends to produce long-living currents (because of the coupling $\beta \rho J$) 
that generate the peak shown in Fig. (\ref{fig:MF}b). 
One can compute the time $\overline{t}$ when $J$ reaches its maximum value
\begin{align} \label{ref:tmax}
    \overline{t} = \frac{1}{\mu} \log \frac{(\alpha + \mu)( \beta - \mu)}{\beta \alpha } \; ,
\end{align}
with $\overline{t}=0$ if $\beta < \alpha + \mu$ (this condition is obtained using the fact that $\overline{t} \geq 0$ in (\ref{ref:tmax})).
Fig. (\ref{fig:MF}c) reports the contour plot $\overline{t}(\alpha,\beta)$ obtained through (\ref{ref:tmax}) with the color indicating the magnitude of $\overline{t}$.
From the equation for $\overline{t}$, one has $\lim_{\beta \to \infty} \overline{t} = \mu^{-1} \log ( 1 + \frac{\mu}{\alpha})$. This asymptotic limit is shown in Fig. (\ref{fig:MF}d).

\begin{figure}[!t]
\centering\includegraphics[width=.45\textwidth]{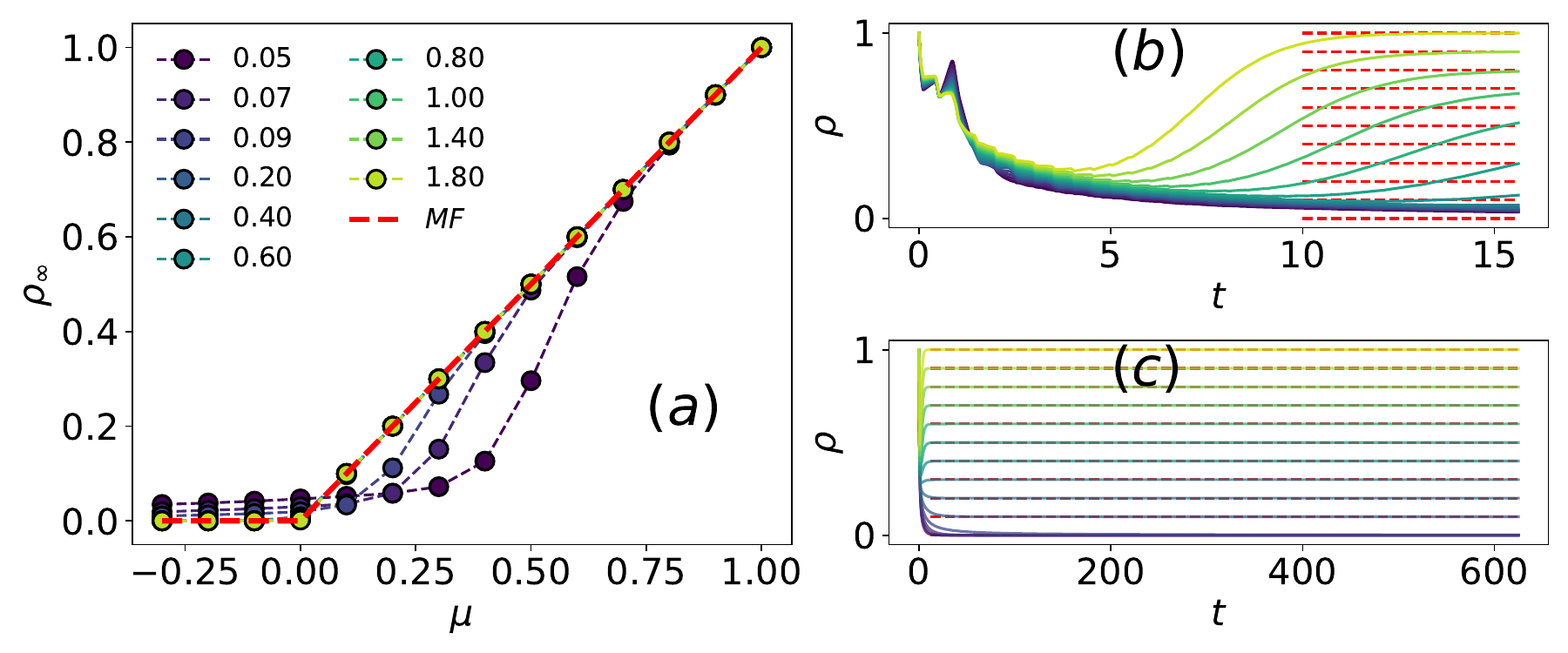}
\caption{Phase Diagram. (a) Stationary density $\rho_\infty$ as a function of $\mu$ for different values of $\alpha$ (increasing values from violet to yellow, see legend). The dashed red line is the mean-field prediction. (b) Time evolution of $\rho(t)$ at $\alpha=0.05$ for different values of $\mu$ (the ones shown in (a), with $mu$ increasing from violet to yellow). (c) $\rho(t)$ at $\alpha=2$ for the same $\mu$ values shown in (b). Dashed red lines represent the mean-field stationary values.}
\label{fig:PD}      
\end{figure}

This happens although the stationary states are the same as the coagulation model in the well-mixed approximation:
\begin{align}
    \lim_{t \to \infty} \rho(t) &\equiv \rho_\infty  = \frac{\mu}{\beta}  \\
    \lim_{t \to \infty} J(t) &\equiv J_\infty = 0 \; ,
\end{align}
as one can also check by setting $\dot{J}=\dot{\rho}=0$ without solving the dynamics. On the other hand, while $J(t)$ depends on $\rho(t)$, at the mean-field level, $\rho(t)$ remains independent of $J(t)$. However, the mean-field computation suggests that non-trivial dynamics might be observed looking, for instance, at $\int dx\, \rho(x,t)$.   
To test the predictions of the mean-field theory, 
Fig. (\ref{fig:PD}a) reports the numerical solution of the equations for $\rho$ and $J$ (here $\rho(t) = \int dx\, \rho(x,t)$ and $J(t) = \int dx \, J(x,t)$). 
\blue{The details on numerical solutions of the equations for $\rho$ and $J$ are provided in the Appendix (\ref{appendix}). As initial conditions, $\rho(x,0)$ is a Gaussian and $J(x,0)=0$. The numerical evolution of $\rho$ and $J$ is performed considering periodic boundary conditions. The initial Gaussian density profile evolves towards a uniform stationary state $\rho_\infty$. The current $J$, initially zero, is nonvanishing on intermediate times and eventually zero in the stationary state.}
The phase diagram has been obtained by varying $\mu$ for different values of the tumbling rate $\alpha$ (here $\beta=1$ and $v=1$). One observes apparent deviations from the mean-field prediction at small $\mu$ as $\alpha$ decreases (and thus the active motion becomes more important).  Looking at the dynamics of $\rho(t)$, one obtains that at small $\alpha$ (panel (b) in Fig. (\ref{fig:PD})) $\rho(t)$ approaches its stationary mean-field value very slowly. On the contrary, as $\alpha$ increases $\rho(t)$ quickly converges to $\rho_\infty$.
One can connect this behavior with the fact that there is a characteristic time $\overline{t}$ which maximizes the current $J$.
In other words, long-living currents slow down the relaxation time of $\rho(t)$ producing an apparent violation of the mean-field prediction.  

\begin{figure*}[!t]
\centering\includegraphics[width=1.\textwidth]{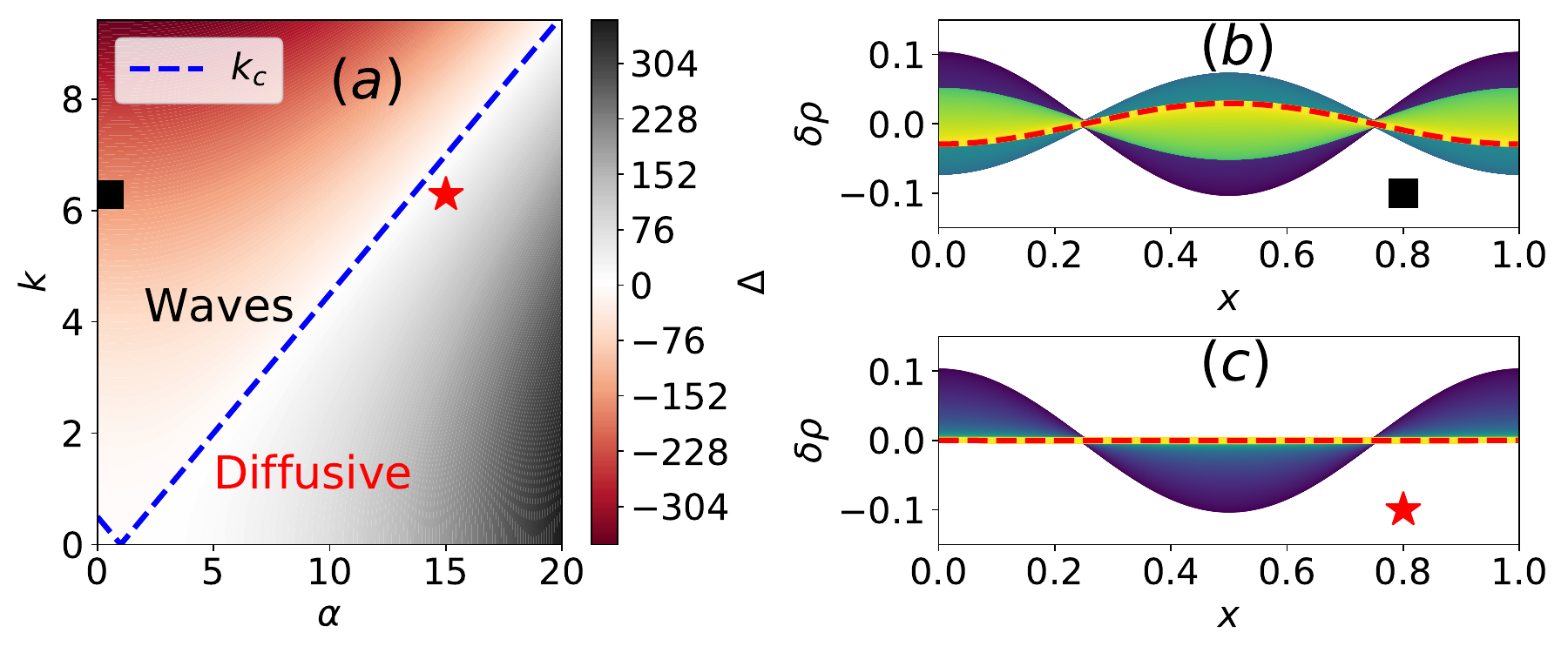}
\caption{Linear stability. (a) Contour plot of $\Delta(\alpha,k)$ for $\beta=v=\mu=1$. The critical wave number $k_c(\alpha)$ is the dashed blue line. For $k>k_c$, before relaxing, a perturbation propagates as a wave, for $k<k_c$ relaxes diffusively to the uniform configuration. (b) Wave-like perturbation for $k=2 \pi$ and $\alpha=0.5$ (see the black square in (a)). (c) Diffusive perturbation for $k=2 \pi$ and $\alpha=15$ (see the red star in (a)). Different curves are taken at different times (time increases from violet to yellow).}
\label{fig:LS}      
\end{figure*}

\subsection*{Linear Stability Analysis}
From the previous section, it follows that the characteristic time scale on which $J$ decays depends on $\rho$ because it is coupled to $J$ through the coupling constant $\beta$. This coupling, at the mean-field level, produces excess in $J$ that decays on a longer time scale (larger than $\alpha^{-1}$). Numerical solutions of the actual dynamics suggest that this effect impact also the dynamics of $\rho$ (that is transparent to $J$ in the mean-field picture).
I now explore the spatiotemporal evolution of the spreading process combined with active motion in finite dimensions through the linear stability analysis of the stationary well-mixed configurations against small perturbations. As a standard procedure, one starts with
\begin{align}
    \rho(x,t) &= \frac{\mu}{\beta} + \delta \rho \\ 
    J(x,t) &= \delta J \; ,
\end{align}
the linearized dynamics for the perturbations $(\delta \rho, \delta J)$ reads
\begin{align}
    \delta \dot{\rho} &= - \delta J^\prime - \mu \delta \rho\\
    \delta \dot{J} &= -v^2 \delta \rho^\prime - \delta J (\alpha + \mu) \; .
\end{align}
Once one introduces $\underline{\phi} \equiv (\delta \rho, \delta J)$, it is possible to rewrite in the compact form
\begin{align}
    \dot{\underline{\phi}} &= A \underline{\phi} \\ 
    A &\equiv \begin{bmatrix}
-\mu & -\partial_x \\
-v^2 \partial_x & -(\alpha + \mu) 
\end{bmatrix} \; .
\end{align}
Looking at damped plane wave solutions $\underline{\phi} \propto \exp{(\sigma(k) t - i k x)}$, the dispersion relation $\sigma(k)$ controls the stability of the perturbation. $\sigma(k)$ is the solution of the eigenvalues equation
\begin{align}
    \det \left[ A - \sigma \mathds{1}\right] = 0
\end{align}
with eigenvalues
\begin{align}
    \sigma(k)_{\pm} 
    =-\frac{\mu + \alpha}{2} \pm \frac{\mu - \alpha}{2} \sqrt{ 1 - \left( \frac{2 v k}{\mu -\alpha}\right)^2 } \; .
\end{align}
The stability condition requires $\Re{ \sigma(k)}>0$ and thus, for $k \to 0$, one obtains $\sigma_+ = -\alpha$ and $\sigma_-=-\mu$, so that the long-wavelength perturbation is always damped ensuring that density fluctuations are diffuse. However, for finite $k$ values, $\sigma(k)$ can acquire an imaginary part so that density fluctuations are dumped oscillations that propagate as waves. This happens whenever
\begin{align}
    \Delta \equiv (\mu - \alpha)^2 - (2 v k)^2 < 0
\end{align}
and thus there is a critical wavelength $k_c$ given by
\begin{align} \label{eq:kc}
    k_c = \frac{\mu - \alpha}{2 v}
\end{align}
that separates purely diffusive excitations for $k<k_c$ from dumped travelling waves for $k>k_c$. From (\ref{eq:kc}) it follows that there is a limit where propagating waves 
dominate up to the macroscopic scale. This happen for $\alpha \to \mu$ so that $k_c \to 0$.
On the other hand, in the large tumbling rate limit $\alpha \to \infty $ there are only diffusive excitations.
The linear stability analysis produces the phase diagram shown in \blue{Fig. (\ref{fig:LS}a)} that has been obtained for $\mu=\beta=v=1$.
To test this prediction one can solve numerically (\ref{eq:rho}) and (\ref{eq:J}) (the initial condition is a small perturbation of the uniform profile $\rho(x,0) = \frac{\mu}{\beta} + \delta_0 \cos(k_0 x)$ with $k_0=2 \pi$ and periodic boundary conditions), with the initial condition on the current $J(x,0)=0$ (with $v=\beta=\mu=1$, $\alpha=0.5,15$). In the case $\alpha=0.5$, one has $k_0 > k_c(\alpha)$ so that linear stability predicts traveling waves. As one can see in \blue{Fig. (\ref{fig:LS}b)}, that shows $\delta \rho(x,t) \equiv \rho(x,t) - \mu / \beta$, the initial plane wave persists in the system while, 
in the second case $\alpha=15$ \blue{(Fig. (\ref{fig:LS}c))}, $k_0 < k_c(\alpha)$, the initial wave dissipate fast into a uniform configuration without oscillations.

\subsection*{Dynamics towards the stationary state}
Although one cannot analytically solve the non-linear dynamics in finite dimensions, it is possible to gain some insight into limiting cases. First, I consider the large $\alpha$ limit.
In this case, there are two possibilities: if the ratio $D_A=v^2 / \alpha$ is maintained finite, i.e., the Brownian limit of the run-and-tumble walker, the model reduces to reaction-diffusion in one spatial dimension. This limiting case can be understood by setting (\ref{eq:J}) in the following form
\begin{align}
    \alpha^{-1} \dot{J} = -D_A \rho^\prime -J [ 1 + \mu / \alpha - \beta \rho / \alpha]
\end{align}
once the limit $\alpha \to \infty$ is performed, one arrives at the following constitutive relation for the current
\begin{align}
    J[\rho] = D_A \rho^\prime 
\end{align}
once this expression of $J$ is inserted into the equation for $\rho$, one arrives at the Fisher-Kolmogorov equation (see for instance \cite{murray2007mathematical}) 
\begin{align} \label{eq:FK-like}
    \dot{\rho} = D_A \rho^{\prime\prime} + \mu \rho \, (1 - \frac{\beta}{\mu} \rho) \; .
\end{align}
Performing the same limit but at finite velocity $v$, the diffusivity drops to zero, i.e., $v \rho^\prime / \alpha \to 0$, and thus one ends with the well-mixed case because $J=0$ so that the dynamics reduces to (\ref{eq:dyn_rhoMF}). In this situation, the spreading process drives the system to uniform configurations $\rho_\infty = \mu / \beta$ so fast that active motion is irrelevant. This is also the case of $\beta \to \infty$ with $v$ finite. There is another limit that is not trivial that I mention but I will not discuss. The case is $\beta \to \infty$ and $v\to\infty$. In this limit, one can introduce a diffusion constant of the spreading process defined as $D_s \equiv v^2 /\beta$. Since $\beta^{-1} J \to 0$, the constitutive relation is $J[\rho] = D_s (\log \rho)^\prime$ that inserted into (\ref{eq:rho}) brings to
the following non-linear diffusion equation
\begin{align} \label{eq:NLD}
    \dot{\rho} = -D_s (\log \rho)^{\prime \prime} +\mu \rho (1 - \frac{\beta}{\mu}\rho) \; .
\end{align}
In the opposite limit, one has $\beta$ as a small parameter 
so that the dynamics of $\rho$ and $J$ is the same as obtained within the linear stability analysis. 

\blue{The numerical integration of the equations for $\rho$ and $J$ is the primary tool for making progress in intermediate regimes.
As suggested by linear stability analysis, one should appreciate some crossover from a situation where the stationary state is approached through dumped traveling waves to another situation where the wave front relaxes following a purely dissipative dynamics. One can move between these two situations by varying the tumbling rate $\alpha$ (keeping the other time scales fixed).}

\begin{figure}[!t]
\centering\includegraphics[width=.45\textwidth]{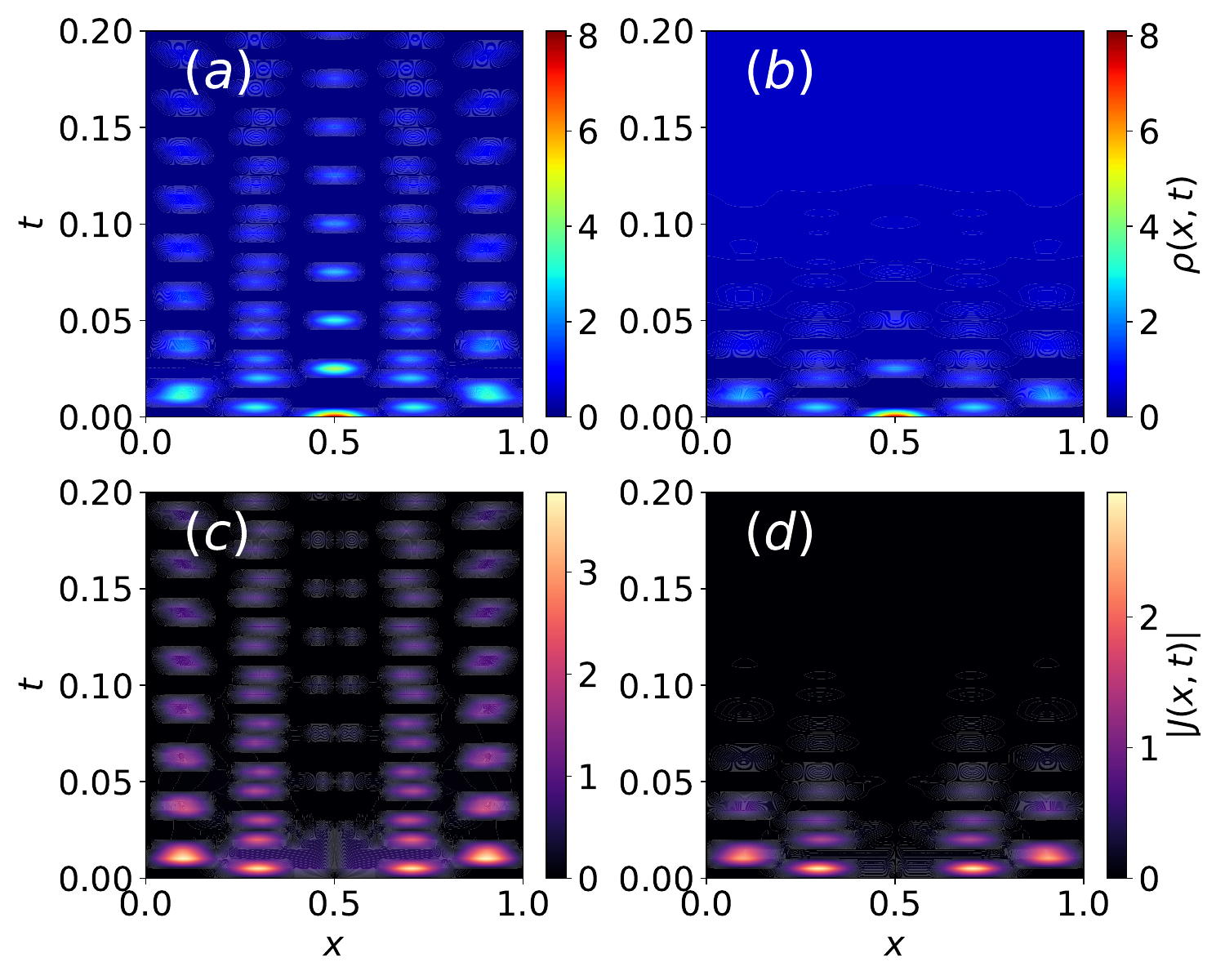}
\caption{Dynamics. The color map indicates the time evolution of the density field $\rho(x,t)$ from an initial Gaussian distribution centered in $L/2$ ($L=1$). Different panels display different values of the tumbling rate $\alpha=0.1,2$ (panels (a) and (b), respectively). Panels (c) and (d) show the intensity of the corresponding current fields $|J(x,t)|$.
\blue{At small $\alpha$ value, the system clearly support traveling damped waves that become more and more damped as $\alpha$ increases.} }
\label{fig:dyn}      
\end{figure}

To test the \blue{existence of these} different scenarios, equations (\ref{eq:rho}) and (\ref{eq:J}) have been solved numerically with a Gaussian profile for $\rho(x,0)$ as the initial condition (while $J(x,0)$ is set to zero).  Fig. (\ref{fig:dyn}) reports the typical time evolution of the density field $\rho(x,t)$ and the current field $J(x,t)$ for two values of tumbling rate ($\alpha=0.5,2$). The reaction parameters are set $\mu=1$ and $\beta=0.5$ so that the stationary state is mixed ($\rho_\infty > 0$). In the case $\alpha=0.5$, the initial Gaussian profile propagates ballistically with a minimal attenuation, this is also mirrored by the behavior of $J(x,t)$ that does not decay. On the contrary, as $\alpha$ increases the traveling wave and currents tend to disappear rapidly. These long-living currents can be interpreted as Fisher waves preventing the system from quickly reaching a stationary uniform state.

\subsection*{Diffusive limit and pattern formation}
In the case of time scales larger than the typical time scale of relaxation of $J(x,t)$, one can \blue{get rid of $J$ and}
write a closed equation for $\rho(x,t)$. 
This is because once one imposes $\dot{J}=0$, it returns a constitutive relation $J=J[\rho]$. I refer to this situation as the diffusive limit.  
The link between $J$ and $\rho$ is
\begin{align}
 J[\rho] &= -D[\rho] \rho^\prime \\ 
 D[\rho] &\equiv \frac{v^2}{\alpha + \mu - \beta \rho} \; .
\end{align}
That brings us to the following diffusive equation
\begin{align} \label{eq:diff_rho}
    \dot{\rho} = \left[D \rho^\prime \right]^\prime + \mu \rho ( 1 - \frac{\beta}{\mu} \rho) \; .
\end{align}
\blue{In this limit, there is a problem related to the behavior of $D[\rho]$ because it becomes negative for large enough $\beta$ values. It is well known that the changing of sign in the diffusion coefficient is a signal of pattern formation \cite{murray2007mathematical}.
Moreover, a negative diffusion coefficient can be regularized by inserting surface terms that are proportional to  $\nabla^4 \rho$ (see, for instance, \cite{cates2010arrested,paoluzzi2020information}). This is because the linear stability analysis of (\ref{eq:diff_rho}) leads to a second-order equation for the wave-length $k$ in Fourier space that defines the dispersion relation $\sigma(k)$. A change of sign of the diffusion coefficient indicates that $\sigma(k)$ changes sign too, and thus an initially small perturbation undergoes an uncontrolled growth. Once we consider the $\nabla^4 \rho$ term, this brings a $k^4$ contribution that in turn stabilizes the emerging patterns in the region of the phase diagram where the diffusion coefficient is negative. However, (\ref{eq:diff_rho}) has a much serious problem because $D[\rho]$, in changing its sign, diverges at $\alpha + \mu - \beta \rho = 0$ whose meaning has to be understood.
We notice that this divergency happens at a critical density $\rho_c = \frac{\alpha + \mu}{\beta} = \frac{\alpha}{\beta} + \rho_{\infty}$. We can get rid of the divergence in the small $\beta$ regime, i.e., assuming  $\mu$ and $\beta $ both small so that $\rho_{\infty}$ is kept fixed. This limit means assuming the spreading process is so slow that variation in the local density due to birth/death processes is so rare that the change in $\rho$ due to them causes a small fluctuation in $J$ that quickly relaxes towards $J=0$. 
}

\begin{figure}[!t]
\centering\includegraphics[width=.45\textwidth]{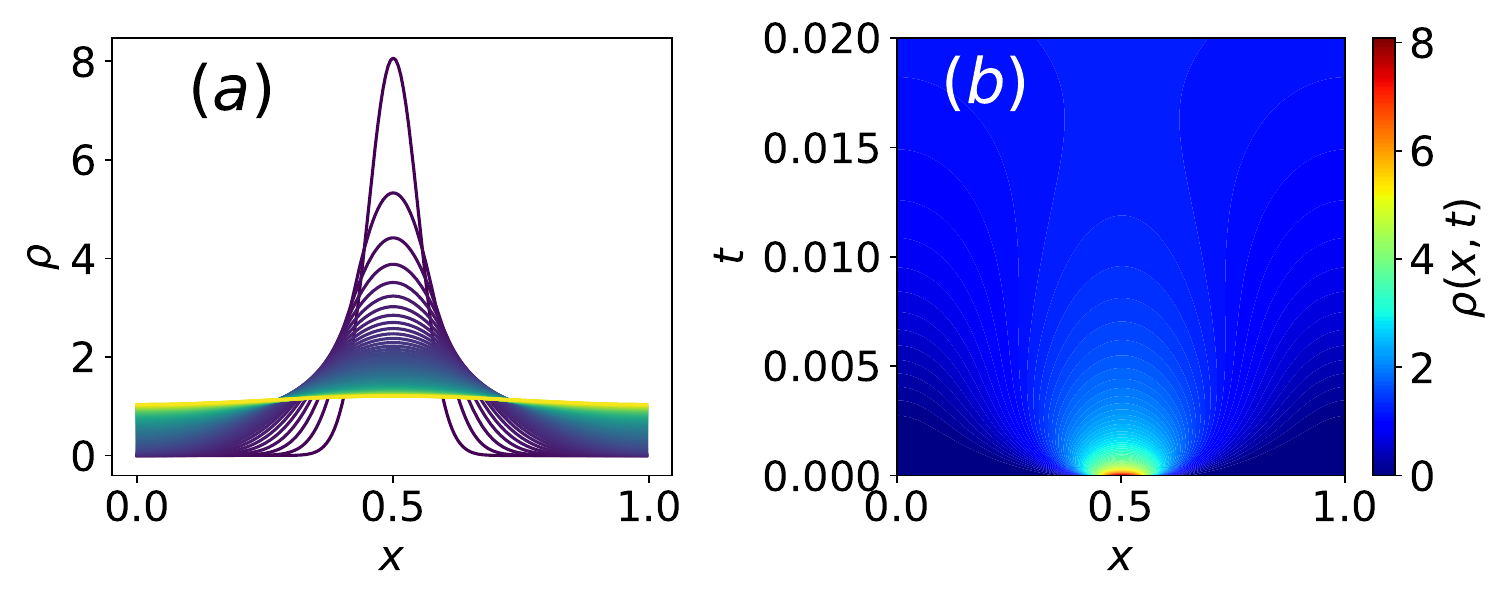}
\caption{Diffusive Limit. (a) Relaxation of a Gaussian density profile towards homogeneous configurations (time increases from violet to yellow). Panel (b) reports the same profile as a color map $\rho(x,t)$.}
\label{fig:diffL}      
\end{figure}

If $\beta$ is small enough, the diffusive limit is well-defined
and thus the diffusion constant is not diverging and remains positive with $D < D_A$. The dynamics is governed by an effective Fisher-Kolmogorov equation whose typical time-evolution is shown in Fig. (\ref{fig:diffL}). 

Going back to (\ref{eq:diff_rho}),
in the case of a perturbation $\delta \rho$ around the uniform stationary state $\rho_{\infty} = \mu / \beta$ (we consider $\mu>0$), the dynamics of $\delta \rho$ is controlled by
\begin{align}
    \delta \dot{\rho} = \left[ D \delta \rho^\prime \right]^\prime - \mu \delta \rho \; .
\end{align}
Again, this equation does not generally make sense unless one considers a small $\beta$ regime so that $D[\rho]$ remains positive and finite. 
In the small $\beta$ limit,
at the zero order in $\delta \rho$, $D[\delta \rho]$
can be replaced by $D\simeq \tilde{D}_A$, with $\tilde{D}_A$ renormalized by the spreading process, i.e., $\tilde{D}_A \equiv v^2 / (\alpha + \mu)$, so that one ends with a diffusive equation for the perturbation 
\begin{align}
    \delta \dot{\rho} =  \tilde{D}_A \rho^{\prime\prime} - \mu \delta \rho \; .
\end{align}
A more interesting situation can be obtained if one keeps the first order in $\delta \rho$ in the expression of $D=D[\rho]$. In particular, considering small perturbation so that $|\delta \rho| < 1 $, upon defining
\begin{align}
    \tilde{D}  &\equiv \frac{D_A}{1 + \frac{\mu}{\alpha}} \\
    \tilde{\beta} &\equiv \frac{\beta}{\beta + \alpha}
\end{align}
for $\tilde{\beta} < 1$, 
one has $D \simeq \tilde{D} \left[1 + \tilde{\beta} \delta \rho \right]$ and thus
\begin{align} \label{eq:kpz_like}
    \delta \dot{\rho} = D[\delta \rho] \delta \rho^{\prime \prime} + \tilde{D} \tilde{\beta} (\delta \rho^\prime)^2 -\mu \delta \rho \; .
\end{align}
Adding a noise term to this equation (representing the effect of the fast degrees of freedom), brings us to
\begin{align} \label{eq:kpz_like2}
    \delta \dot{\rho} = D[\delta \rho] \delta \rho^{\prime \prime} + \tilde{D} \tilde{\beta} (\delta \rho^\prime)^2 -\mu \delta \rho + \eta
\end{align}
with $\eta \equiv \eta(x,t)$, $\langle \eta(x,t) \rangle = 0$, and $\langle \eta(x,t) \eta(y,s) \rangle =2 T \delta(x-y) \delta(t - s)$ (with $T$ setting the strength of the noise). Considering fluctuations due to the noise around the absorbing state critical point $\mu=0$, it follows that (\ref{eq:kpz_like2}) is formally equivalent to Kardar-Parisi-Zhang equation (KPZ) model \cite{PhysRevLett.56.889} (but with a non-linear diffusion coefficient). This mapping suggests that pattern formation at criticality should fall into the KPZ universality class. It is interesting to note that the strength of the KPZ term depends on the ratio $\mu/\alpha$, and thus on both autonomous motion and replication.

\section{Conclusions and Outlook} 
This work addresses a simple one-dimensional continuum model that combines two non-equilibrium phenomena: active dynamics and absorbing state phase transitions. 
\blue{Wave-front propagation in biological systems is a typical example of the application of the model introduced here.}
The continuum model consists of a gas of run-and-tumble particles 
that interacts via a reaction process that interacts via a reaction process.  This process annihilates particles at a rate $\beta$ and produces new particles at a rate $\mu$. 
Although the presence of active motion does not change the stationary properties of the system, the dynamics towards the absorbing or mixed state qualitatively changes with $\alpha$ from a situation where the system fast relaxes because of diffusive modes, to another situation, for small $\alpha$, where the system can support propagating waves. 
Consequently, the relaxation time towards the stationary state strongly depends on the interplay of active motion with the spreading process. 
\blue{It is possible to rationalize this interplay already at level of a men-field approximation where $\rho(t)$ does not vary in space. In. mean-field,}
while $\rho(t)$ follows the standard logistic evolution, the current $J(t)$ does not decay monotonously \blue{because of its coupling with $\rho(t)$: interaction that is generated }  
by the coagulation dynamics. In particular, $J(t)$ admits a local maximum for large $\beta$. This prediction agrees with the behavior of integrated quantities such as  $A(t) = \int dx \, A(x,t)$, with $A=\rho,J$. These quantities were computed numerically beyond the mean-field regime by solving the full dynamics. 
Spatial heterogeneities couple currents and density, causing $\rho(t)$ to relax very slowly to $\rho_\infty$ at small $\alpha$. This dynamical slowing down is distinct from the usual critical slowing down typical of critical phenomena, having a genuinely non-equilibrium origin. 
The density-current coupling generates damped waves that can propagate up to macroscopic scales or rapidly diffuse, depending on the interplay between the spreading process and active motion.

Considering the system's dynamics on time scales where the current is stationary (i.e., $\do{J}=0$), I also demonstrated that, in general, it is not safe to consider the diffusive limit of the model due to the competition between the spreading dynamics and active motion. 
This is manifested by a pathological diffusive limit that leads to diverging and eventually \blue{a} negative diffusion constant. 
On phenomenological grounds, the negative diffusion constant in the diffusive continuum model indicates pattern formation and can be addressed by introducing appropriate surface tension terms. 
\blue{The model admits a well-defined diffusive limit in the small $\beta$ limit}, with fluctuation dynamics governed by a KPZ-like equation.

The independence of the stationary state from non-equilibrium active dynamics may be a model-dependent feature of the continuum description considered here. Like its equilibrium counterpart, the active coagulation model belongs to the directed percolation universality class. Future work should investigate the effects of activity near the critical point, using both lattice models and continuum descriptions, to determine whether activity alters the morphology of pattern formation.

\subsection*{Acknowledgments}
M.P. acknowledges funding from the Italian Ministero dell’Università e della Ricerca under the programme PRIN 2022 ("re-ranking of the final lists"), number 2022KWTEB7, cup B53C24006470006.

\appendix
\section{ \blue{Numerical Solution of the Dynamics}} \label{appendix}

\blue{The equations for $\rho$ and $J$ given by (\ref{eq:rho}) have been solved numerically using the Euler scheme for the time integration.
The numerical implementation consists in mesh the space in $N_{grid}$ space interval of size $\Delta x= 1/N_{grid}$ (with periodic boundary conditions) 
and the time in $N_t$ intervals of size $\Delta t$.
The value of the fields on the lattice point $i=1,...,N_{grid}$ at time $t + \delta t$ are updated as follows
\begin{align} \label{eq:dis}
\rho^i_{t+\delta t} &= \rho^i_t + f^i_\rho \Delta t \\ 
f^i_\rho &\equiv -\nabla J_i^t + \rho^i_t (\mu - \beta \rho^i_t) \\
J^i_{t+\delta t} &= J^i_t + f^i_J \Delta t \\ 
f^i_J &\equiv -v^2 \nabla\rho_i - J^i_t \left( \alpha + \mu - \beta \rho ^i_t \right)
\end{align}
where the gradients $\nabla g_i$ of a generic function $g_i$ defined on the mesh have been computed using the central finite difference method
\begin{align}
\nabla g_i \equiv \frac{g_{i+1} - g_{i-1}}{2 \Delta x} \; .
\end{align}
The initial condition is a Gaussian profile for $\rho(x,0)$ with zero current $J(x,0)=0$. 
The time spacing is fixed by the Courant–Friedrichs–Lewy condition $\Delta t \leq \frac{1}{2} \frac{\Delta x^2}{D_a}$, with $D_a\equiv v^2/\alpha$. 
In the case of (\ref{eq:dis}) $\Delta t  = \frac{1}{2} \frac{\Delta^2}{D_a}$ with $N_{grid}=400$, and $N_t = 10^7$.
The dynamics in the diffusive limit (\ref{eq:diff_rho}) has been solved in the same way
\begin{align}
\rho^{i}_{t + \delta t} &= \rho^i_t + h_\rho^i \Delta t \\ 
h_\rho^i &\equiv \nabla \left[ \mathcal{D}_i \rho_i^t \right] + \rho_i (\mu - \beta \rho_i^t) \\ 
\mathcal{D}_i &\equiv \frac{v^2}{\alpha + \mu - \beta \rho_i^t} \nabla \rho_i^t 
\end{align}
with $N_{grid}=200$, $N_t=2\times 10^6$, and $\Delta t = \frac{1}{10} \frac{\Delta x^2}{D_a}$.
}
\bibliography{bib}
\bibliographystyle{rsc}

\end{document}